\documentclass[manuscript]{aastex63}

\newcommand{\rsun}{$R_\sun$}
\newcommand{\kms}{km s$^{-1}$}

\usepackage{bm}
\usepackage{graphicx}
\usepackage{subfigure}

\begin{document}
 
\title{Formation of a Magnetic Cloud from the Merging of Two Successive Coronal Mass Ejections}

\author{Chong Chen}
\affiliation{School of Microelectronics and Physics, Hunan University of Technology and Business, Changsha 410205, People's Republic of China; \href{mailto:chenc@hutb.edu.cn}{chenc@hutb.edu.cn}}

\author{Ying D. Liu}
\affiliation{State Key Laboratory of Space Weather, National Space Science Center, Chinese Academy of Sciences, Beijing 100190, People’s Republic of China; \href{mailto:liuxying@swl.ac.cn}{liuxying@swl.ac.cn}}
\affiliation{University of Chinese Academy of Sciences, Beijing 100049, People’s Republic of China}

\author{Bei Zhu}
\affiliation{Space Engineering University, Beijing 101416, People’s Republic of China}

\author{Huidong Hu}
\affiliation{State Key Laboratory of Space Weather, National Space Science Center, Chinese Academy of Sciences, Beijing 100190, People’s Republic of China; \href{mailto:liuxying@swl.ac.cn}{liuxying@swl.ac.cn}}

\author{Rui Wang}
\affiliation{State Key Laboratory of Space Weather, National Space Science Center, Chinese Academy of Sciences, Beijing 100190, People’s Republic of China; \href{mailto:liuxying@swl.ac.cn}{liuxying@swl.ac.cn}}

\begin{abstract}

On 2022 March 28 two successive coronal mass ejections (CMEs) were observed 
by multiple spacecraft and resulted in a magnetic cloud (MC) at 1 AU.
We investigate the propagation and interaction properties of the two CMEs 
correlated with the MC using coordinated multi--point remote sensing 
and in situ observations from Solar Orbiter, STEREO A, SOHO, and Wind. 
The first CME was triggered by a filament eruption with a high inclination angle.
Roughly 9 hr later, the second CME originating from the same active region erupted 
with a smaller tilt angle and faster speed compared to the first one.
The second CME overtook the preceding CME and formed a merged front 
at approximately 75 \rsun{}, which developed into a complex ejecta at 1 AU.
The descending speed and low proton temperature inside the complex ejecta suggest that 
the two CMEs have fully merged before reaching 1 AU, 
leading them to begin expanding rather than compressing against each other.
The complex ejecta appears to have the magnetic field and plasma signatures of an MC,
although there is a discontinuity in the magnetic field implying previous interactions.
The cross section of the complex ejecta, 
reconstructed from in situ data using a Grad-Shafranov technique, 
exhibits a right--handed flux rope structure.
These results highlight that an MC--like complex ejecta lacking interaction features could arise from the complete merging of two CMEs.

\end{abstract}

\keywords{Solar coronal mass ejections (310) --- Solar terrestrial relation (1473) --- Solar wind (1534)}

\section{Introduction} \label{sec:intro}
Coronal mass ejections (CMEs) are massive magnetized plasma structures ejected from the solar atmosphere.
They may interact with each other when propagating from the corona through interplanetary space (e.g., \citealt{Gopalswamy2001ApJ, Gopalswamy2002GeoRL, Manoharan2004JGRA, Farrugia2004AnGeo}).
CME-CME interactions are associated with magnetic reconnection, momentum exchange,
and changes in the CME geometries (see \citealt{Lugaz2017SoPh} and references therein).
The interactions between CMEs can lead to difficulties in interpreting the magnetic structure observed at 1 AU and make it difficult to predict space weather.

A diverse range of structures can emerge at 1 AU as a consequence of the complex interactions between CMEs.
One such structure is the complex ejecta defined by \citet{Burlaga2001JGR, Burlaga2002JGRA}.
Compared to typical isolated magnetic clouds (MCs), a complex ejecta may or may not have organized magnetic structures at 1 AU (e.g., \citealt{Burlaga1987JGR, Burlaga2001JGR, Burlaga2002JGRA, Vandas1997JGR, Wang2003JGRA1081370W, Farrugia2004AnGeo}).
A fast CME with a shock interacting with another preceding CME could produce two major complex structures at 1 AU: a shock inside an interplanetary coronal mass ejection (ICME; e.g., \citealt{Burlaga1987JGR, Vandas1997JGR, Liu2014ApJ793L41L, Lugaz2015GeoRL424694L, Lugaz2015JGRA1202409L, Liu2018ApJ859L4L, Xu2019ApJ884L30X}) and ``ICME-in-sheath'' where an ICME is embedded inside a sheath region \citep{Liu2020ApJ897L11L}. They both have high potential to trigger intense geomagnetic storms due to the enhancements in the magnetic field, speed, and density (e.g., \citealt{Burlaga1987JGR, Tsurutani1997GMS, Liu2014NatCo, Meng2019JGRA}).
A long-duration structure with smooth rotational magnetic field vectors but not a typical MC can also arise from CME-CME interactions. 
\citet{Dasso2009JGRA} examine the formation of the 2005 May 15 ICME, which lasted about 2 days at 1 AU, based on in situ measurements, remote sensing observations, and radio emission.
They find that this long-duration event was likely to be associated with two interacting MCs of nearly perpendicular orientations. 
\citet{Lugaz2013ApJ} perform a simulation of two CMEs with near-perpendicular orientations and find that the consequence is a prolonged structure with a smooth rotation in the magnetic field vectors. 
By combining these simulation results and the analysis of 2001 March 19–22 long-duration event, \citet{Lugaz2014GeoRL} find that such event might seem like a single isolated structure and appear to have no clear features of interactions. However, the magnetic field vectors manifest a smooth rotation over an extended duration and the size of the ejecta is twice larger than a typical MC. 
\citet{Liu2014ApJ793L41L} investigate the plasma and magnetic field characteristics of two interacting CMEs near 1 AU and suggest that after fully merging they may form an MC-like structure without clear signs of CME interactions.
These varying structures at 1 AU denote the diverse processes of CME-CME interactions and their different impacts on the magnetosphere.
Therefore, understanding how CME-CME interaction changes ICME in situ signatures 
can deepen our understanding of the CME collisions and their geomagnetic effects.

Combining remote sensing observations and in situ measurements from multiple spacecraft allows to track the interactions between CMEs from the Sun to 1 AU and investigate how a complex ejecta is formed.
The links between in situ signatures and the CME interaction features in images can also be investigated.
For example, with wide-angle heliospheric imaging observations from the Solar Terrestrial Relations Observatory (STEREO; \citealt{Kaiser2008}) and in situ measurements from multiple points,
\citet{Liu2012ApJ746L15L} study the evolution of three interacting CMEs from the Sun all the way to the Earth. 
The launch of the Solar Orbiter (SolO; \citealt{Muller2013SoPh}) makes it possible to observe CMEs 
from a unique vantage point and perform joint observations with other spacecraft.
The Full Sun Imager (FSI) telescope, which is the wide field channel of the Extreme Ultraviolet Imager (EUI, \citealt{Rochus2020AA}) onboard SolO, has much wider field of view (FOV) than that of any previous solar extreme ultraviolet (EUV) imager.
FSI can provide information about the early evolution of CMEs in the corona and play a critical connecting role between solar disk features and structures captured by coronagraphs.
The Multi Element Telescope for Imaging and Spectroscopy (METIS; \citealt{Antonucci2020AA}) is an externally
occulted coronagraph onboard SolO that simultaneously images the full off-limb solar corona in both visible and ultraviolet light in a square FOV of $\pm$2.9\degr{} width.
With its capability to observe the global corona, METIS can trace the early propagation and evolution of CMEs, thus enhancing our understanding of CME initiation and dynamics.
The Solar Orbiter Heliospheric Imager (SoloHI; \citealt{Howard2020AA}) is 
a next-generation heliospheric imager, which is composed of four 2048$\times$1920 pixel tiles.
It possesses an FOV spanning approximately 5\degr{}-45\degr{} to the east of the Sun,
achieving an effective resolution comparable to the LASCO/C2 coronagraph onboard Solar and Heliospheric Observatory (SOHO; \citealt{Domingo1995}) during the SolO perihelion.
Therefore, with coordinated remote sensing and in situ observations from SolO and other spacecraft, the capability of continuously tracking CMEs in the Sun-Earth space and providing detailed observations of CMEs is being further enhanced (e.g., \citealt{Bemporad2022AA665A7B, Hess2023AA679A149H, Niembro2023FrASS1091294N, Liberatore2023ApJ957110L}).

On 2022 March 28, two CMEs erupted from NOAA AR 12975 (N12\degr{}W01\degr{}), which were associated with two M-class flares that peaked at about 11:29 UT and 20:59 UT respectively. They resulted in an MC at 1 AU, which is unusual.
We analyze the observations from the remote sensing and
in situ instruments on board SolO, STEREO A, SOHO, and Wind to trace the formation of the MC.
The results are helpful for understanding the connections between in situ ejecta and CME interaction features.

\section{observations and analysis} \label{source}
\begin{figure} 
\begin{interactive}{animation}{movie1.mp4}
	\centering
	
	\begin{minipage}{0.5\textwidth}
		\centering
		\includegraphics[width=\textwidth]{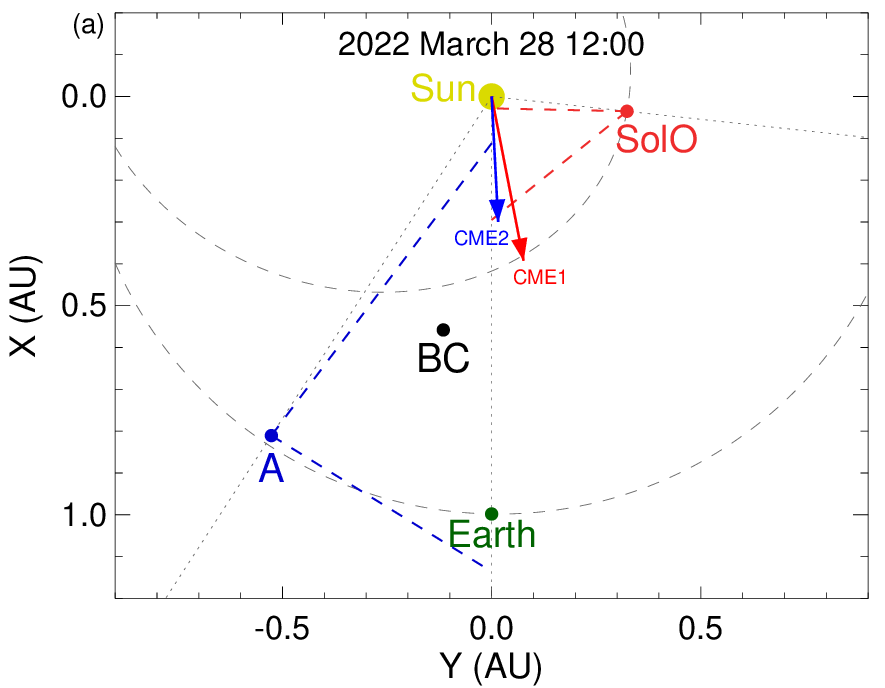}  
	\end{minipage}
	\begin{minipage}{0.86\textwidth}
		\centering
		\includegraphics[width=\textwidth]{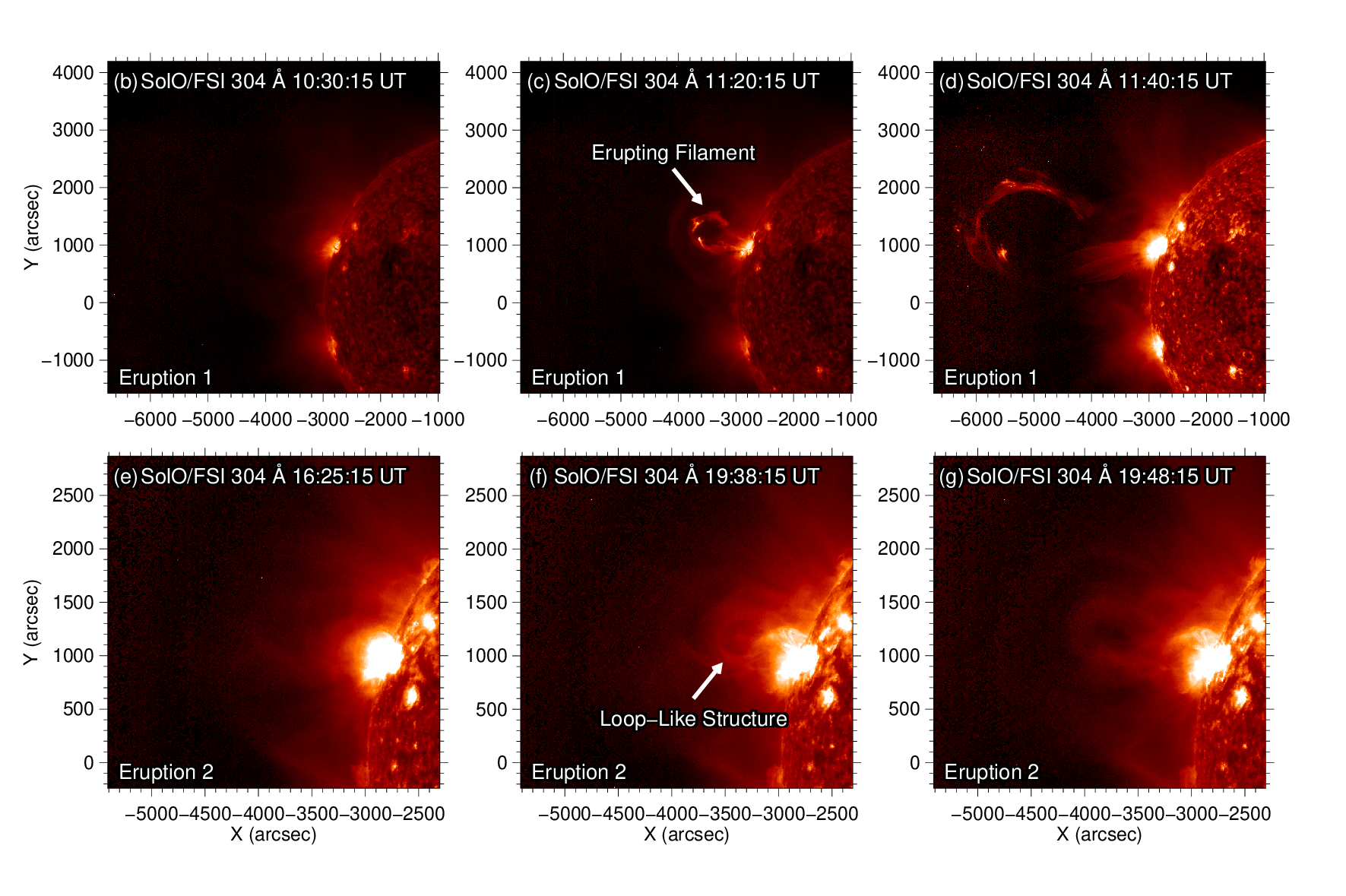} 
	\end{minipage}
\end{interactive}
	\caption{\label{f1}(a) Positions of the spacecraft and planets in the ecliptic plane on 2022 March 28. The ``A'', ``SolO'', and ``BC'' represent the satellites of STEREO A, Solar Orbiter, and BepiColombo respectively. The blue and red arrows indicate the propagation directions of the CMEs. The blue and red dashed lines coming from the spacecraft show the FOVs of STEREO A/HI and SoloHI on 2022 March 28, respectively. (b)-(g) Enhanced EUV images of the two CMEs viewed from SolO/FSI at 304 \AA{} showing the two CME eruptions. Panels (b)-(d) are accompanied by an animation that displays the first CME eruption process from 10:30 UT to 12:40 UT with a duration of 3 s. Panels (e)-(g) are accompanied by the same animation that displays the second CME eruption process from 16:15 UT to 20:58 UT. The animation shows the first/second CME eruption at the left/right side on each frame of the video. The arrows are removed in the animation.\\
	(An animation of this figure is available.)}
	
\end{figure}

The positions of the spacecraft in the ecliptic plane on 2022 March 28 are shown in Figure \ref{f1}(a).
STEREO A and SolO are 0.97 and 0.32 AU from the Sun, and 33\degr{} east and 84\degr{} west of the
Earth, respectively.
This configuration of SolO enables the side-view observation of Earth-directed CMEs and establishes a favorable common viewing geometry with STEREO A.
According to the CME propagation directions (marked by the red and blue arrows in Figure \ref{f1}(a)),
the two CMEs may be imaged from three different points including STEREO A, SOHO, and SolO and detected by BepiColombo \citep{Benkhoff2021SSRv21790B} and Wind.
SolO/FSI 304 \AA{} enhanced images are displayed in panels (b)–(g). 
The half-FOV of FSI is about 2 \rsun{} during the time of eruptions, 
and the CMEs are limb events as viewed from SolO, so we can track the CME eruptions at the early stage
using EUV observations. 
The first CME (CME1) occurred on about 2022 March 28 10:30 UT, caused by a filament eruption with a peak speed of about 700 \kms{} (panels (b)–(d)). 
A possible associated feature on the Sun is an M4.0 flare from NOAA AR 12975 (N12\degr{}W01\degr{}).
The axis of the filament possibly has a largely inclined angle, 
because the two legs separate with a large distance.
After about 9 hr, the second CME (CME2) associated with
an M1.0 flare from the same active region erupted at about 19:38 UT on March 28 (panels (e)–(g)).
Despite the faint signature of the erupting CME2 in EUV images, it attained an estimated peak speed of about 1000 \kms{}, faster than that of the preceding CME1.
As denoted by the white arrow in panel (f), CME2 appeared to a loop-like structure at the 
eruption stage indicative of a small tilt angle.
The two CMEs are likely to interact each other in the heliosphere, given to their propagation directions (Figure \ref{f1}(a)) and launch times on the Sun. 

\begin{figure}
\begin{interactive}{animation}{movie2.mp4}
	\epsscale{1}
	\centering
	\resizebox{0.75\textwidth}{!}{
		\plotone{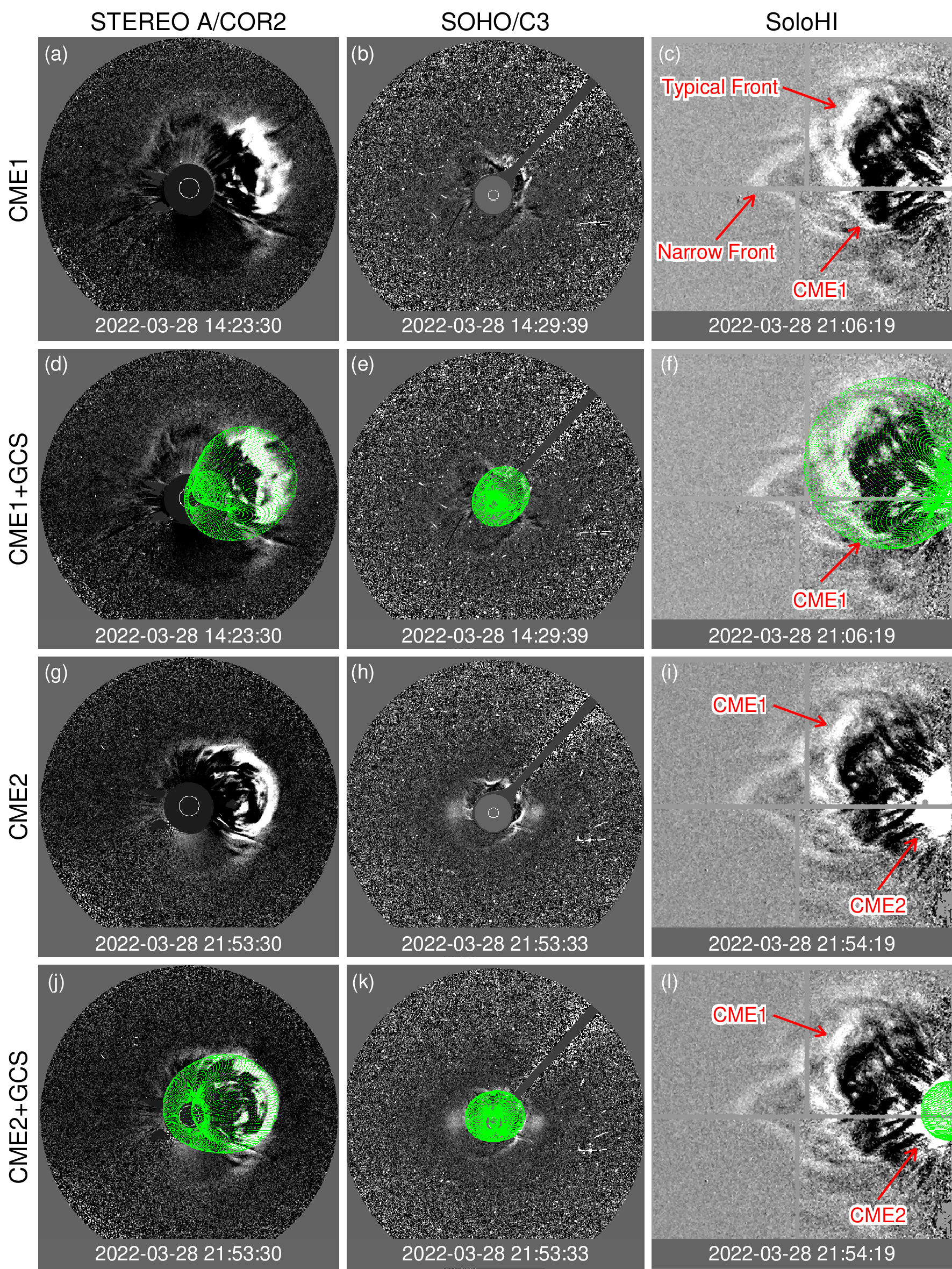}
	}
\end{interactive}
	\caption{\label{f2}Running-difference white-light images and corresponding GCS modeling (green grids) of the two CMEs from STEREO A/COR2 (left), SOHO/C3 (middle), and SoloHI (right). Detectors and times are given in the images. The times given in the SoloHI images correspond to the average observation time of the four tiles. Panels (b) and (h) are accompanied by an animation that displays the two CME propagation processes in the FOVs of SOHO/LASCO C2 and C3 from 2022 March 28 09:51 UT to 2022 March 29 05:29 UT. Panels (c) and (i) are accompanied by the same animation that displays the two CME propagation processes in the FOV of SoloHI from 2022 March 28 01:30 UT to 2022 March 29 21:10 UT. The animation shows the LASCO images at the left side and the SoloHI images at the right side on each frame of the video. The arrows are removed in the animation.\\
	(An animation of this figure is available.)} 
\end{figure}

Running-difference white-light images of the two CMEs from STEREO A/COR2, SOHO/LASCO, and SoloHI
are shown in Figure \ref{f2}.
SoloHI images are a composite of four tiles, which are separated by the narrow rectangular boundaries crossing the whole image. 
During the events, SolO is around the perihelion with an FOV of roughly 6-50 \rsun{} at the west limb of the Sun.
Therefore, the CME propagates from the right to the left of the SoloHI image.
Around the SolO perihelion, SoloHI has a cadence of 12 minutes on the inner two tiles (the right column in the SoloHI image) and 24 minutes on the outer tiles (the left column in the SoloHI image).
Therefore, there are slight misalignments between the four tile images,
and we use the average time of the four tiles to represent the observation time.
CME1 appeared in the FOVs of STEREO A/COR2 at about 11:53 UT, LASCO/C2 at about 12:12 UT, and SoloHI at about 13:30 UT.
For CME1, the observation times of the SoloHI images do not match closely with those of the COR2 and LASCO images, therefore, the timestamp of the SoloHI image in panels (a)-(f) is different from the times of COR2 and LASCO observations.
CME1 appeared as a full halo CME in the LASCO coronagraphs, which indicates that CME1 propagated toward to the Earth between SolO and STEREO A.
There are two distinct bright fronts as marked by the red arrows in the SoloHI image.
The `narrow' front seems to be independent of the typical front \citep{Hess2023AA679A149H}.
Therefore, we analyse the geometry and the kinematics of CME1 based on the typical front.
CME2 was also a halo CME and appeared in the FOVs of STEREO A/COR2 at about 20:23 UT, LASCO/C2 at about 20:36 UT, and SoloHI at about 21:54 UT.
It was imaged by three different points almost simultaneously at about 21:53 UT (panels (g)-(i)).
The SoloHI image indicates that CME2 was chasing CME1 from behind given their eruption speeds and launch times on the Sun.

Based on the three different points of view 
provided by STEREO A/COR2, SOHO/LASCO, and SoloHI,
we use a graduated cylindrical shell (GCS) model 
proposed by \citet{2006Thernisien} to determine 
the CME geometry and trajectory near the Sun.
The GCS model can determine the direction of propagation, 
tilt angle of the CME flux rope, and height (e.g., \citealt{2009Thernisien, 2010bLiu, chen2019ApJ, 2017Hu}). 
By adjusting the parameters, 
the GCS model fits the two CMEs from 
the three viewpoints very well (see Figure \ref{f2}).
This highlights the good complementarity and 
synergy of the SoloHI observations with other datasets.
The GCS modeling results show that CME1 propagates 
approximately 11\degr{} westward and 10\degr{} northward 
relative to the Sun-Earth line and the ecliptic plane, 
while CME2 propagates about 3\degr{} westward and 6\degr{} northward.
The propagation directions of the two CMEs are generally similar 
and consistent with the location of the active region (N12\degr{}W01\degr{}).
The tilt angles of the two CME flux ropes obtained from the GCS model are about 50\degr{} 
and 6\degr{} with respect to the ecliptic plane, 
which are consistent with the FSI 304 \AA{} observations (see Figure \ref{f1}(b)-(g)).
CME2 had a higher speed than CME1, so CME2 was catching up with CME1.

\begin{figure}
\begin{interactive}{animation}{movie3.mp4}
	\centering
\begin{minipage}[b]{0.48\linewidth}
	\raisebox{0.15cm}{\includegraphics[width=\linewidth]{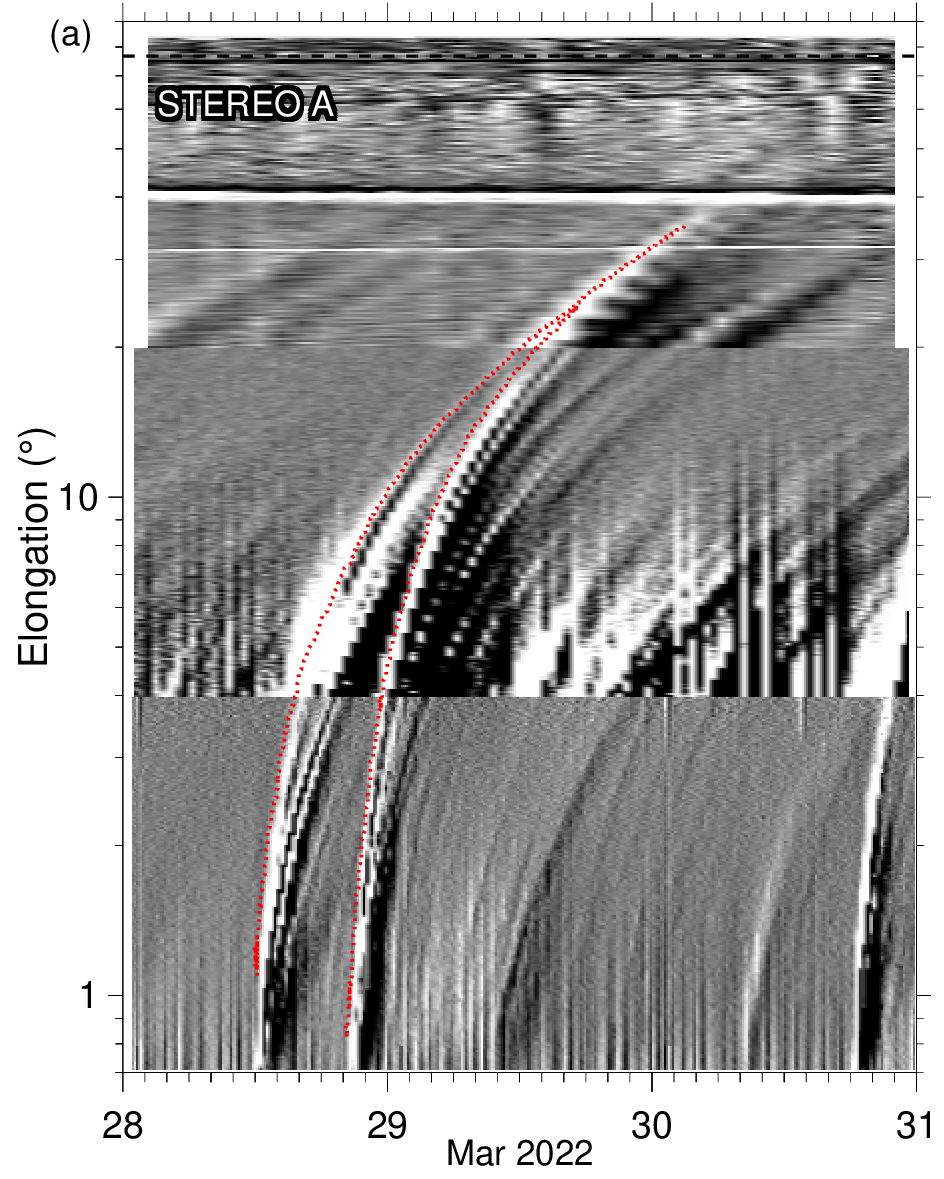}} 
\end{minipage}
	\begin{minipage}[b]{0.49\textwidth}
		\includegraphics[width=\linewidth]{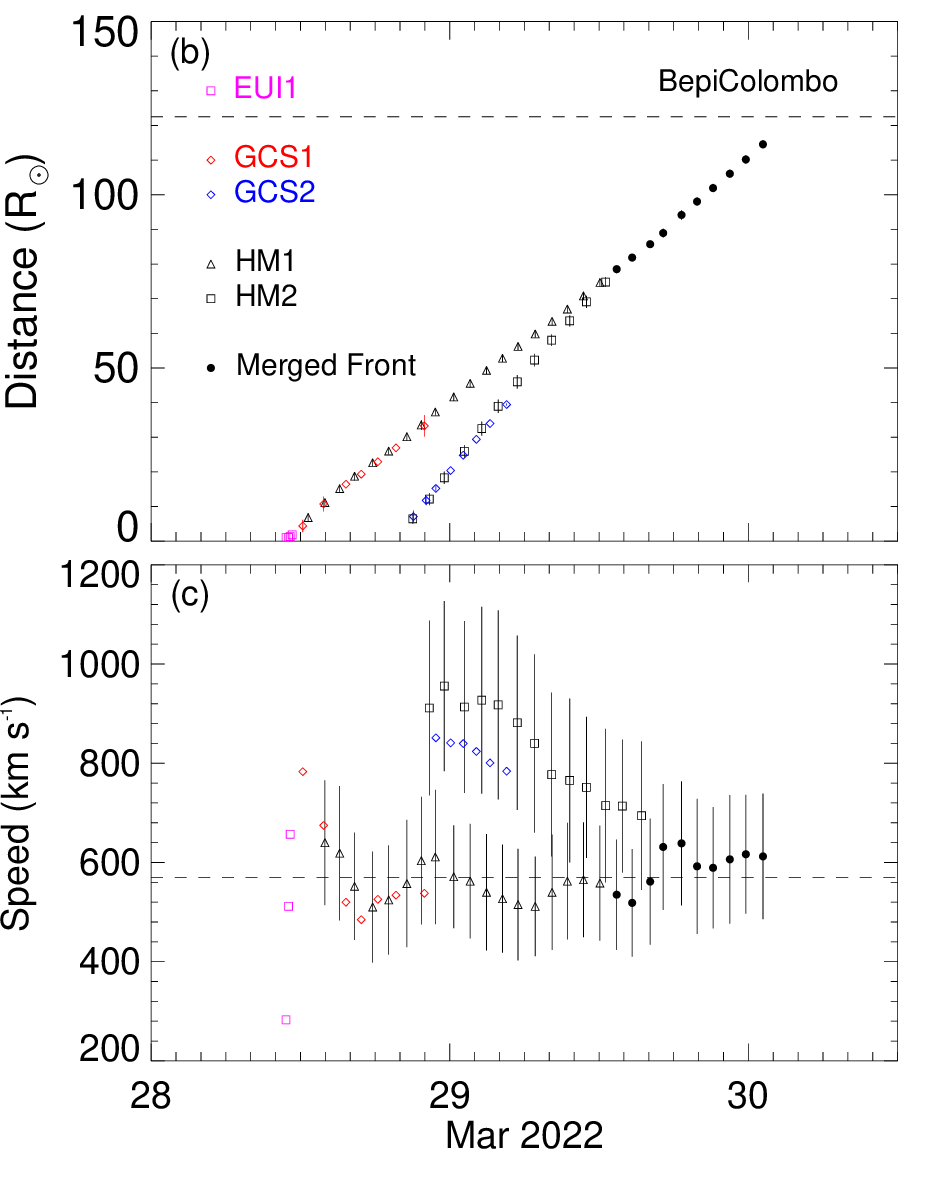}  
	\end{minipage}  
\end{interactive}
	\caption{\label{f3}(a) Time-elongation map constructed from running-difference images of COR2, HI1 and HI2 from STEREO A along the ecliptic plane. The red curves indicate the tracks of the CMEs, along which the elongation angles are extracted. The black dashed horizontal line denotes the elongation angle of the Earth. (b)-(c) Radial distance and speed profiles of the CME leading edges derived from EUV observations, the GCS model and the HM approximation. Note that the filled circle points represent the distance and speed of the merged CME front. The horizontal dash line in the panel (b) indicates BepiColombo's heliocentric distance of about 122 \rsun{}. The horizontal dashed line in the panel (c) represents the shock speed (about 570 \kms{}) at Wind. Panel (a) is accompanied by an animation that displays the two CME propagation processes in the FOVs of COR2, HI1 and HI2. The animation starts at 2022 March 28 00:23 UT and ends at 2022 April 1 22:08 UT with a duration of 12 s. \\
	(An animation of this figure is available.)}

\end{figure}

We investigate the evolution of the two CMEs 
in interplanetary space by producing a time-elongation map 
(e.g., \citealt{2008Sheeley, 2009Davies, 2010aLiu, 2010bLiu}).
Since SolO was largely observing the CMEs from behind (see Figure \ref{f1}(a)), 
it is not feasible to jointly investigate the CME evolution with STEREO A. 
Moreover, the time-elongation map based on SoloHI contains long-time data gaps. 
As a result, we only use white light running-difference images from STEREO A to trace the evolution of the two CMEs.
The time-elongation map displayed in the panel (a) of Figure \ref{f3} is produced by stacking
running-difference images of COR2, HI1, and HI2 from STEREO A within a slit along the ecliptic plane.
The tracks associated with the two CMEs are denoted by the red curves in the map.
Note that the CME1 `narrow' front is not visible in the map, and we use the typical front of CME1 to analyse its kinematics, as mentioned above.
As illustrated in the map, the tracks associated with CME1 and CME2 appear to converge eventually, 
revealing the impending merging between the two CMEs before reaching 1 AU.
The elongation angles of the two CME leading edges in the ecliptic plane are extracted along the tracks, 
which can then be converted into radial distances based on the methods summarized in \citet{2010bLiu}.
We derive the CME kinematics employing the harmonic mean (HM) approximation proposed by \citet{2009Lugaz}, which assumes that CMEs propagate as a spherical front attached to the Sun and along a fixed radial direction.
We use the propagation directions from GCS modeling in converting elongations to distances.

The height and speed of the two CMEs obtained from the GCS modeling, HM approximation,and EUV observations
are displayed in the panels (b)-(c) of Figure \ref{f3}.
Because just three EUV images are available from SolO/FSI for CME2, its distance and velocity
derived from EUV observations are not provided.
As shown in the panel (b), for both CMEs, the heights derived from different methods can be connected consistently.
The horizontal dashed line represents the location of BepiColombo, 
which heliocentric distance is about 122 \rsun{}.
We can see that CME2 caught up with CME1 and merged into a single CME front at approximately 75 \rsun{}. 
Therefore, BepiColombo could measure the complex structure resulting from the CME interactions earlier than Wind.
In situ data from BepiColombo will be useful to investigate the newly merged structure, and its comparison with the in situ data from Wind will tell how the complex structure evolves from BepiColombo to Wind.
We plan to do an analysis of the BepiColumbo in situ data in a forthcoming study.
The speeds of the two CMEs are shown in the panel (c).
CME1 was quickly accelerated to about 800 \kms{}, 
then showed a rapid deceleration, and finally moved with a nearly constant speed.
This is an intermediate-speed CME speed profile: a quick acceleration, then a rapid deceleration, and finally a nearly constant speed \citep{Liu2016ApJS}.
CME2 gradually decelerated from approximately 980 \kms{} and then propagated with a constant value of approximately 600 \kms{} after merging with CME1.
The merged CME front will reach Earth at approximately 09:15 UT on March 31 with a speed of about 600 \kms{}, estimated from the linear fit to the distances of the merged front (i.e., the distances after 75 \rsun{}).

\begin{figure}
	\epsscale{1}
	\plotone{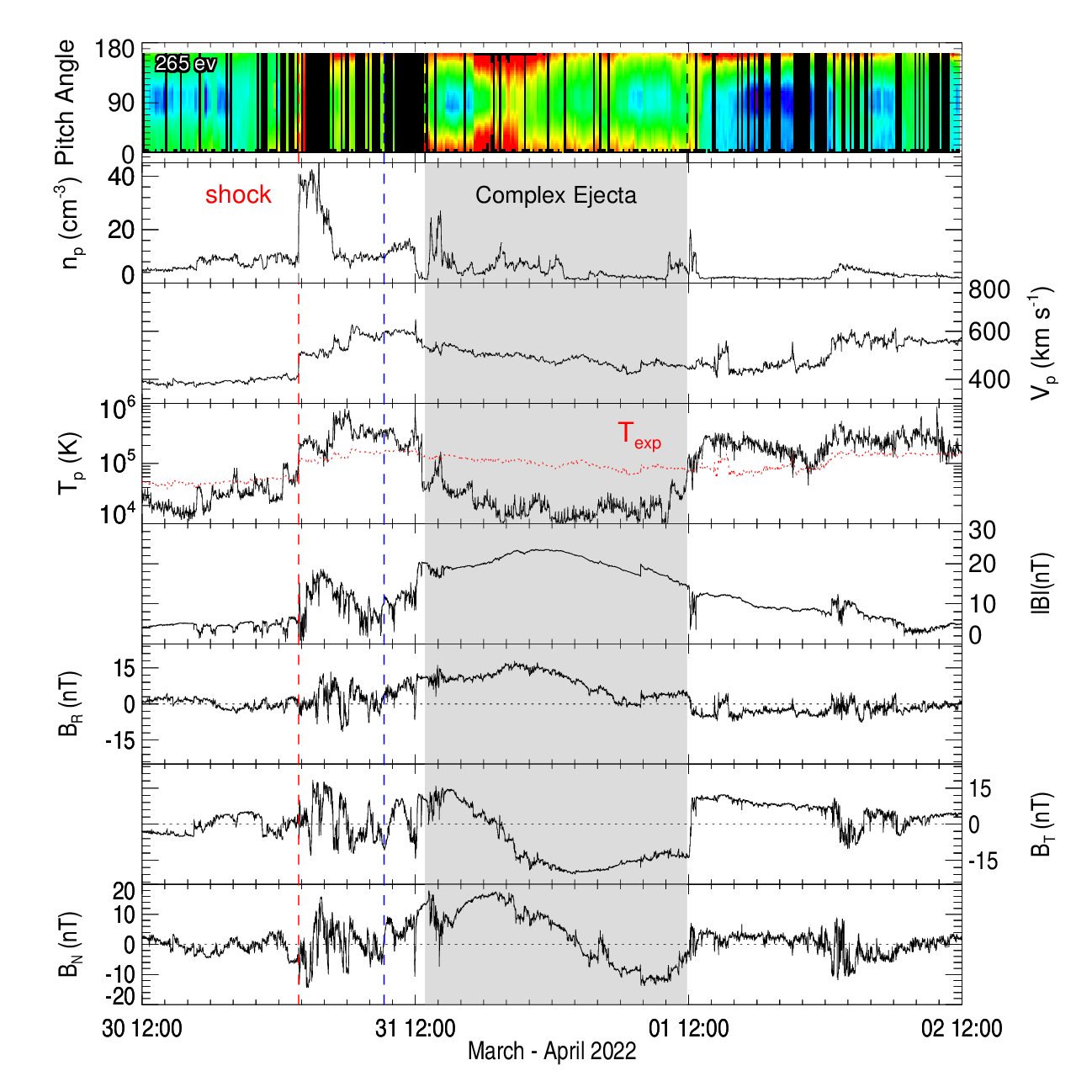}
	\caption{\label{f4}Solar wind plasma and magnetic field parameters at Wind associated with the 2022 March 28 CMEs. From top to bottom, the panels show the pitch angle distribution of 265 eV electrons, proton density, bulk speed, proton temperature, and magnetic field strength and components, respectively. The dotted line in the fourth panel denotes the expected proton temperature calculated from the observed speed \citep{1987Lopez, Richardson1995JGR}. The shaded region indicates the complex ejecta interval determined by the GS reconstruction. The red vertical dashed line denotes the observed arrival time of the shock. The blue vertical line represents the predicted arrival time of the merged front at the Earth.} 
\end{figure}

Figure \ref{f4} shows the in situ measurements at Wind. 
A shock is observed at Wind around 01:45 UT on March 31, followed by a sheath region, which is from 01:45 UT to 12:51 UT on March 31 lasting about 11 hr. 
The individual CME boundaries cannot be distinguished within the complex ejecta.
The magnetic field and plasma signatures, such as the bidirectional streaming electrons (BDE), 
the descending speed profile, the low proton temperature, and strong magnetic fields, 
indicate that the two CMEs have merged into a single ICME.
Further more, the descending speed and low proton temperature within the complex ejecta 
suggest that the two CMEs have fully merged before arriving at 1 AU, 
leading them to start expanding instead of compressing against each other.
The complex ejecta appears to have the signatures of an MC, 
i.e., a smooth and strong magnetic field, a coherent rotation of the field, 
and a low proton temperature, according to the definition of \citeauthor{1981Burlaga} (\citeyear{1981Burlaga}; also see \citealt{2008Echer, Meng2019JGRA} and references therein).
Therefore, we use a boundary-sensitive reconstruction technique (see below) 
to determine the interval of the complex ejecta.
The interval is from 12:51 UT on March 31 to 11:52 UT on April 1, which is roughly consistent
with the interval identifed based on the signatures of the BDE, temperature and magnetic field.
The arrival time (09:15 UT on March 31) at the Earth predicted by wide-angle imaging
observations is in good agreement with the sheath. 
The shock speed is about 570 \kms{}, 
which is also roughly consistent with the predicted speed ($\sim$600 \kms{}).
Although the complex ejecta seems to be an isolated expanding MC, 
there is a discontinuity around 08:00 UT on April 01 which may indicate previous CME interactions. 

\begin{figure}
	\epsscale{1}
	\plotone{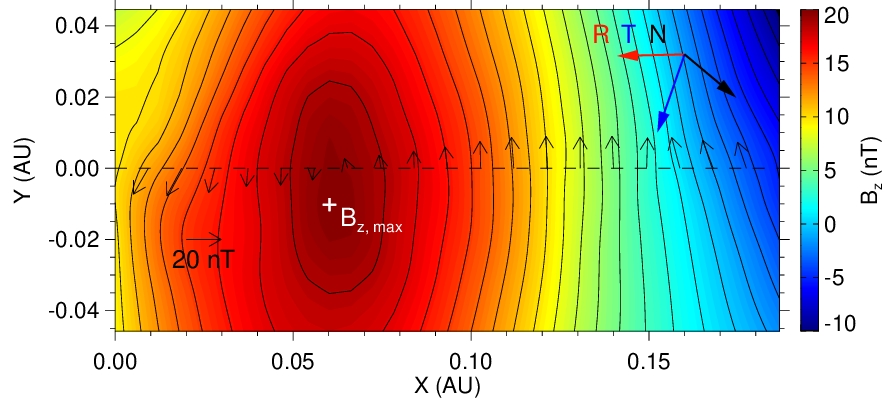}
	\caption{\label{f5}Reconstructed cross section of the ICME at Wind. Black contours are the distribution of the vector potential, and the color shading shows the value of the axial magnetic field strength. The location of the maximum axial field is indicated by the white cross. The horizontal dash line marks the trajectory of the Wind spacecraft. The thin black arrows denote the direction and magnitude of the observed magnetic field projected onto the cross section, and the thick colored arrows show the projected \textbf{RTN} directions.} 
\end{figure}

The cross section of the complex ejecta reconstructed from the in situ data 
using a Grad-Shafranov (GS) technique \citep{1999Hau, 2002Hu} is displayed in Figure \ref{f5},
which is helpful to understand the in situ signatures caused by the CME merging.
The reconstruction results give a right-handed flux rope structure, 
which can be obtained from the transverse fields along the trajectory of Wind.  
As Wind moves along the horizontal dashed line from the left to the right (positive $x$ direction), 
it would observe a $B_{R}$ component that is largely positive,
a $B_{T}$ changing from positive to negative, 
and a $B_{N}$ changing from positive to negative. 
The elevation angle of the flux rope is about 41\degr{} and the azimuthal angle is about 330\degr{}.
The flux rope tilt angle is similar to that of CME1 (50\degr{}) determined from the GCS model near the Sun.
Therefore, we conjecture that the tilt angle of CME1 remains largely unaffected by the later impact from CME2.

There are two similar events resulting from the interaction of two CMEs that have been discussed before \citep{Lugaz2014GeoRL,Liu2014ApJ793L41L}.
Based on a simulation of the interaction of two CMEs with near-perpendicular orientations and the analysis of 2001 March 19-22 complex event, \citet{Lugaz2014GeoRL} find that the resulting complex ejecta exhibits characteristics of a single event such as the monotonic declining velocity profile and smooth magnetic field strength. However, there is an interface in the magnetic field indicating the interaction. 
This is similar to our results, suggesting the CMEs have undergone complete merging and 
started to expand after merging.
However, the MC in our work results from two successive CMEs with orientations about 45\degr{} apart. 
This suggests that an MC can also arise from two
interacting CMEs with non-perpendicular orientations.
An interacting and merging complex ejecta is reported by \citet{Liu2014ApJ793L41L}, in which the proton temperature is not depressed as usual for an ICME at 1 AU and the speed profile is not declining monotonically, indicating the merging of two CMEs is still ongoing. 
They suggest that once the CME interactions are completed, the complex ejecta would resemble an MC-like structure without clear signs of CME interactions.
Although the structure in \citet{Liu2014ApJ793L41L} does not display the characteristics of an MC, the individual CME boundaries already cannot be distinguished.

\section{conclusions} \label{sum}
We have investigated the propagation and interaction properties of two 
successive CMEs in relation to the formation of an MC at 1 AU,
using remote sensing observations from SolO, STEREO A and SOHO and in situ measurements at Wind.
The two CMEs, which occurred on 2022 March 28, erupted from the same active region 
with their orientations about 45\degr{} apart. 
The first CME has an intermediate-speed event speed profile: a quick acceleration, then a rapid deceleration,
and finally a nearly constant speed, while the second CME has a gradual deceleration phase followed by a nearly invariant speed.
The second CME overtook the first CME at approximately 75 \rsun{},
and formed a complex ejecta with MC signatures: the BDE, the descending speed profile, 
the low proton temperature, and rotating and enhanced magnetic fields.
These characteristics suggest that the two CMEs have fully merged before arriving at 1 AU, and started expanding after merging.
The reconstructed cross section of the complex ejecta using the GS technique 
indicates a right-handed flux rope configuration, 
whose interval is roughly consistent with the interval identified from the in situ signatures.
Although the complex ejecta seems to be an isolated expanding MC, 
there may be a discontinuity indicating past interactions.
The reconstructed flux rope tilt angle matches the orientation 
of the first erupted CME derived from the GCS model near the Sun. 
This suggests that the later impact from the second CME did not 
greatly alter the tilt angle of the first CME.
These results highlight that an MC-like complex ejecta can arise 
from the complete merging of two CMEs.
Only with wide-angle imaging observations from multiple spacecraft 
can we track and determine its origin from the interaction and merging of two different CMEs.

\acknowledgments
The research was supported by the Strategic Priority Research Program of the Chinese Academy of Sciences (NO. XDB0560000), the NSFC (grants 42274201, 42150105, and 12073032), the National Key R\&D Program of China (No. 2021YFA0718600 and No. 2022YFF0503800), the Research Foundation of Education Bureau of Hunan Province (No. 23B0593), and the Specialized Research Fund for State Key Laboratories of China. 
Solar Orbiter is a space mission of international collaboration between ESA and NASA, operated by ESA. The EUI instrument was built by CSL, IAS, MPS, MSSL/UCL, PMOD/WRC, ROB, LCF/IO with funding from the Belgian Federal Science Policy Office (BELSPO/PRODEX PEA 4000112292 and 4000134088); the Centre National d’Etudes Spatiales (CNES); the UK Space Agency (UKSA); the Bundesministerium für Wirtschaft und Energie (BMWi) through the Deutsches Zentrum für Luft-und Raumfahrt (DLR); and the Swiss Space Office (SSO). The Solar Orbiter Heliospheric Imager (SoloHI) instrument was designed, built, and is now operated by the US Naval Research Laboratory with the support of the NASA Heliophysics Division, Solar Orbiter Collaboration Office under DPR NNG09EK11I. We acknowledge the use of data from STEREO, SOHO, and Wind.

\clearpage

\clearpage

\clearpage

\clearpage

\clearpage

\clearpage
\bibliography{article}
\bibliographystyle{aasjournal}

\end{document}